# Terahertz radiation-induced sub-cycle field electron emission across a split-gap dipole antenna

Jingdi Zhang[1,2]*, Xiaoguang Zhao[3]*, Kebin Fan[3], Xiaoning Wang[3], Gu-Feng Zhang[1], Kun Geng[2],
Xin Zhang[3]†, Richard D. Averitt[1,2]†

1. Department of Physics, University of California, San Diego, La Jolla, CA 92093, USA
2. Department of Physics, Boston University, Boston, MA 02215, USA
3. Department of Mechanical Engineering, Boston University, Boston, MA 02215, USA

*These authors contribute equally to this work. †Corresponding author. E-mail: xinz@bu.edu, raveritt@ucsd.edu

Abstract:

We use intense terahertz pulses to excite the resonant mode (0.6THz) of a micro-fabricated dipole antenna with a vacuum gap. The dipole antenna structure enhances the peak amplitude of the in-gap THz electric field by a factor of ~170. Above an in-gap E-field threshold amplitude of ~10 MVcm$^{-1}$, THz-induced field electron emission is observed (TIFEE) as indicated by the field-induced electric current across the dipole antenna gap. Field emission occurs within a fraction of the driving THz period. Our analysis of the current (*I*) and incident electric field (E) is in agreement with a Millikan-Lauritsen analysis where *log (I)* exhibits a linear dependence on *1/E*. Numerical estimates indicate that the electrons are accelerated to a value of approximately one tenth of the speed of light.

In the last decade, the rapid development of technologies for high efficiency generation of THz waves has enabled the study of strong field-matter interactions in the THz (i.e. far-infrared) region of the electromagnetic spectrum[1,2,3]. Progress in studying non-equilibrium dynamics initiated by intense THz fields [4] includes coherent spin waves[5], the Higgs-mode in superconductors[6] and nonlinear carrier dynamics in semiconductors[7,8,9]. Through leveraging the large local field enhancement arising from E-field confinement within the capacitive gap of metamaterial resonators, a local THz field of several MVcm$^{-1}$ initiated the insulator-to-metal transition in $VO_2$[10] and impact ionization in GaAs metamaterials[11]. Furthermore, given the strongly enhanced local electric field, the metallic resonator-vacuum interface may also be a candidate for coherent electron source. Electron emission and coherent manipulation were demonstrated by exciting metallic nano-size tip at shorter wavelength (near-IR and mid-IR) [12,13,14]. In the THz regime, electron field emission induced ionization of air has been observed [15,16], and the air-plasma based THz source induced electron emission spectrum has also been measured[17].

In this letter, we report the observation of a dynamic process where the locally enhanced THz electric field in the gap of a dipole antenna results in electron field emission into vacuum. This results from field-driven tunneling ionization. This tunneling ionization process, induced by a sub-cycle optical field, was predicted for near-IR excitation of metallic nano-structures[14]. With the development of high-field THz sources, this naturally extends to THz frequencies, where multiphoton ionization and above-threshold ionization are negligible due to the low frequency of the THz pulses.

The experimental concept is depicted in Fig. 1a, where THz pulses impinging on Au dipolar structures induce field emission. To generate THz pulses of sufficient field strength, we use near-IR pulses (800nm, 1.5mJ, 35 fs) from a 1 kHz Ti:sapphire regenerative amplifier laser for tilted-pulse-front generation of intense THz pulses in $LiNbO_3$[1]. After propagating through a grating (groove density 1800 nm$^{-1}$) and cylindrical lenses in a 4-f configuration (Fig. 1b), the wavefront of the incident near-IR pulse is tilted by 62º inside the $LiNbO_3$ crystal to phase-match the group velocity of the near-IR and THz pulses. The output THz beam is collimated and tightly focused on the sample plane with a peak amplitude of ~200 kV/cm. The THz pulse is measured in the time-domain with electro-optical (EO) sampling[18].

The gold field emission electric dipole resonators (Fig. 1c) are fabricated on a 500 nm thick insulating $SiN_x$ film by conventional photolithography[19]. A square shaped region of the $SiN_x$ substrate in the vicinity of the dipole antenna gap is etched away, such that this hollow gap will allow the motion of free electrons that are emitted across the antenna gap. Figure 1d shows the transmission as a function of frequency for these structures. There is a distinct resonance at 0.6 THz corresponding to the dipole resonance of these structures.

Finite-difference time-domain simulations (CST Microwave studio) of the in-gap field enhancement in the time domain and frequency domain are shown in Fig. 2a and b, respectively. The field enhancement has a strong spatial dependence (Fig. 2c, d) and is approximately one order of magnitude larger than a resonator structure where the gaps are filled with semiconductor or dielectric material[10][11][20]. The spectral amplitude peaks at a resonance frequency 0.6 THz. At this frequency, an enhancement ranging from 150 to 400 is achieved across the edge. Fig. 2c, d show numerical simulations of the spatial distribution of the instantaneous in-gap field enhancement at the peak of the time-domain field. Across the capacitive gap, the spatial distribution (Fig. 2d) of field enhancement factor peaks at the antenna-vaccum interface, reaching a value of 170, exponentially decaying (decay length ~ 0.1 µm) to a constant value of 60 in the middle of the gap. The in-gap THz electric field can induce electron emission and, in turn, accelerate the electrons.

To detect electrons generated by the in-gap THz field, the device was placed into a vacuum chamber (cryostat) under high vacuum of $10^{-6}$ mbar. Two ends of the dipole antenna are wired to a circuit, which measures the voltage change across a load resistor of 50 kΩ. The output signal is fed to a lock-in amplifier, synchronized to the 500 Hz modulation frequency of the THz beam (Fig. 1b).

We characterized two identical samples with the setup depicted in Fig. 1b. Figure 3a shows the measured current from the dipole antenna array with respect to the peak E-field amplitude of the incident THz pulse. The current from two samples (blue dots and red squares) show the same threshold of 60kV/cm (free space E-field amplitude), and the nice overlap of the two curves indicates that the THz-induced electron emission (TIFEE) is a *repeatable* effect. To verify the current is from the dipole antenna array, the sample is rotated by 90 degrees to avoid the resonance mode (i.e. the dipole antenna is now orthogonal to the E-field of THz pulse). If the signal arises from the antenna gap no current signal is expected for THz excitation in this orientation. Indeed, no current was detected when the E-field is perpendicular to the antenna. The E-field dependence of the current (*I*) and the existence of a 60 kV/cm threshold unambiguously reveal that the field electrons have been generated by the enhanced in-gap THz field. The corresponding threshold value of in-gap electric field is estimated to be 10.2 MV/cm, taking into account the field enhancement factor of 170. Strikingly, it is close to the reported emission threshold of 10 MV/cm under a static electric field[21].

To analyze the THz-induced field electron emission (TIFEE) data, we use Fowler-Nordheim (F-N) tunneling[21], which involves the wave nature of electrons and Fermi-Dirac statistics for electrons. F-N tunneling predicts that the electron emission *I-V* characteristic is given by an empirical form: $I = CV^{\kappa}\exp(-B/V)$, where *C, B, κ* are fitting parameters for the experimental data[22]. It yields an exact straight line when the *I-V* plot takes the form log(*I/V$^\kappa$*) versus *1/V*. As a special case (*κ*=0), *I* exhibits an exponential dependence on the applied voltage *V*, i.e. a plot log(*I*) vs 1/*V* yields a straight line. Such a semi-log plot is referred to as a Millikan-Lauritsen plot (M-L plot)[23][24]. Indeed, Figure 3b shows that our TIFEE data is in agreement with a linear dependence of *log (I)* and *1/E$_{THz}$*. Although M-L plots are generally used to describe the field emission from materials biased by a static E-field, the reasonable agreement of our data with the fit indicates that the responsible mechanism is F-N-like tunneling.

Insight into the THz field-induced tunneling can be obtained in terms of the Keldysh parameter[25] $\gamma$, which provides criteria for classifying different types of electron emission.

In particular, $\gamma = \omega \frac{\sqrt{2m_e \Phi}}{efE}$, where $\omega$ is the frequency of the THz pulse (though are pulses are broadband, we use 0.6 THz which is the resonant frequency of our structure), $m_e$ is the electron rest mass, $\Phi$ is the work function (5.5 eV for gold), $e$ is the electron charge, $f$ is the field enhancement factor in frequency domain (~400 near the gap edge at 0.6 THz, Fig. 2b) due to the dipole antenna, and $E$ is the peak amplitude (threshold value 60 kV/cm) of the incident THz pulse. For the present experiments, this yields a Keldysh parameter of 0.012. This is consistent with tunneling ionization, since $\gamma \ll 1$ [25].

Tracking the emitted electrons in detail is beyond the reach of the current experiment setup. However, a spatial adiabaticity parameter ($\delta = \frac{l_F}{l_q}$) analysis [12][13][14] can offer a glimpse into the electron motion in a qualitative way. The quiver amplitude is given by $l_q = efE/m_e\omega^2$ and $l_F = 0.1$ μm is the near-field decay length (see Figure 2d). For our experimental configuration, the quiver amplitude is estimated to vary from 12.7 to 42.5 μm, corresponding to incident THz fields of 60 kV/cm to 200 kV/cm, respectively. Thus, the quiver amplitude is much larger than both the near-field decay length $l_F$ and the 2 μm dipole gap, implying that in-gap electrons will not have enough time or space to quiver before they cross the gap. That is, forward motion is favored over quiver motion. Hence, we can neglect the contribution of the Ponderomotive force, which results in the asymmetric average quiver motion of electrons in an oscillating electric field with a strong gradient. The fact that δ and γ≪1 implies that the majority of the ionized electrons finish the trip across the 2 μm gap within a fraction of an oscillating THz period meaning that the electron emission is a sub-cycle process (see supplemental material). Recent experiments using infrared pulses for electron emission from a gold nano-tip shows that the divergence of the electron bunch from the nano-tip can be suppressed by using laser pulses of longer wavelength, which favors sub-cycle emission and quenches the quiver motion of the electrons. Therefore, TIFEE is an extreme example of the complete suppression of quiver motion, due to the local field enhancement, and may work as a well-collimated electron source.

Figure 4 shows SEM images of dipole elements with and without exposure (~1 hr) to the THz field. Figure 4a, b are images of the pristine antenna structure (capacitive gap ~ 2 μm). Figure 4,c, d show the images of the modified gold antenna after exposure. Near the antenna gap, the amorphous gold has moved inwards, shrinking the gap size to 1 μm, as indicated by Figure 4d. The morphology consists of nanometer size gold rods that grow inwards. The orientation and the size of the gold nano-rods is highly ordered, and nominally lines up with the local THz electric field lines. This suggests that the responsible mechanism is due to either collision of high-energy field emitted electron across the dipole antenna gap, where the excessive kinetic energy melt the gold tips, or electromigration caused by strong electric currents driven by the intense THz electric field inside the conductive gold antenna. In the immediate vicinity of the gap, we attribute the "annealed" near-gap gold nano-rods to the melting of the gold structure induced by the impact of the field emitted electrons. In principle, electromigration could also contribute. However, a recent experiment[16] demonstrating air breakdown induced by enhanced THz fields (in a split-ring resonator (SRR) structure) shows that in-gap dipole fusing only occurs at low air pressures (~mTorr). Interestingly, dipole fusing is absent at ambient pressure, while the deformation of at the corners of the SRRs (where the field is also strongly enhanced) still occurs. This suggests that the gold deformation occuring at the SRR corners results from electromigration, while the in-gap nano-rod formation is due to the direct bombardment by high-energy field emitted electrons. Further, this is consistent with the fact that fusing is not observed under ambient pressure. Indeed, similar tungsten nano-structures have been observed due to the high energy helium plasma irradiation[26][27].

In conclusion, we have measured THz induced field electron emission current from split-gap dipole antennas, which is Millikan-Lauritsen-like. Our resonant structure provides strong local field enhancement of the incident light at the designed THz frequency, which efficiently generates ultrafast electron pulses in a sub-cycle regime, free of quiver motion. It will be of great interest to investigate the energy spectrum of the electrons and to demonstrate coherent control of TIFEE in future experiments.

**Acknowledgement:** We acknowledge support from DOE - Basic Energy Sciences under Grant No. DE-FG02-09ER46643, under which the THz measurements and electron acceleration analysis were performed. In addition we acknowledge the National Science Foundation under Grant No. ECCS-1309835. The authors would like to thank Boston University Photonics Center for technical support.

**Figure caption:**

Figure 1. (Color online) (a) Schematic of THz induced field electron emission. A near single cycle THz pulse drives the resonance mode of dipole antenna. At the antenna gap, electrons (blue spheres) escape from the gold-vacuum interface, generating a transient current signal. (b) Experimental setup consisting of the high field THz spectrometer and the field emitted electron detection circuit. The electric field amplitude of the THz pulses are controlled with a pair of wire grid polarizers (WGP). A Si wafer is used to filter out the residual near-IR (800nm) pulses. ZnTe, Wollaston prism (WP) and balanced diode (BD) for electro-optical sampling. (c) Microscope image of the dipole antenna array. The antenna gap is 2 μm. Black squares are regions where $SiN_x$ has been etched to avoid in-gap electric contact. (d) Experimental field transmission spectrum of the dipole antenna array with resonance at 0.6THz.

Figure 2 (Color online) (a), (b) show the time-domain and frequency-domain in-gap field enhancement of the incident THz pulse, probed at the center and the edge of the dipole antenna gap. (c) Simulated two-dimensional spatial distribution of field enhancement. (d) Field enhancement factor to the time-domain THz pulse across the gap (blue dots, simulation result; solid curve, exponential fit). The maximum field enhancement of 170 times is achieved at the gap edge. The decay length of enhanced field near the tip is around 0.1 μm.

Figure 3 (Color online) (a) Current signal due to electron emission as a function of the incident THz pulse peak amplitude from sample A (blue circles) and B (red squares). The electron emission shows a threshold at an incident field strength of 60kV/cm. (b) Millikan-Lauritsen plot of the transient current vs. inverse of the incident E-field (log *I vs. 1/E*). Data from both samples yield a straight line.

Figure 4 (a, b) SEM image of a pristine dipole antenna gap. Long term (>1hr) exposure to the high field THz radiation restructures the dipole antenna shape. (c) (d) show the reformed gold antenna. The zoom in reveals the fine structure inside the gap.

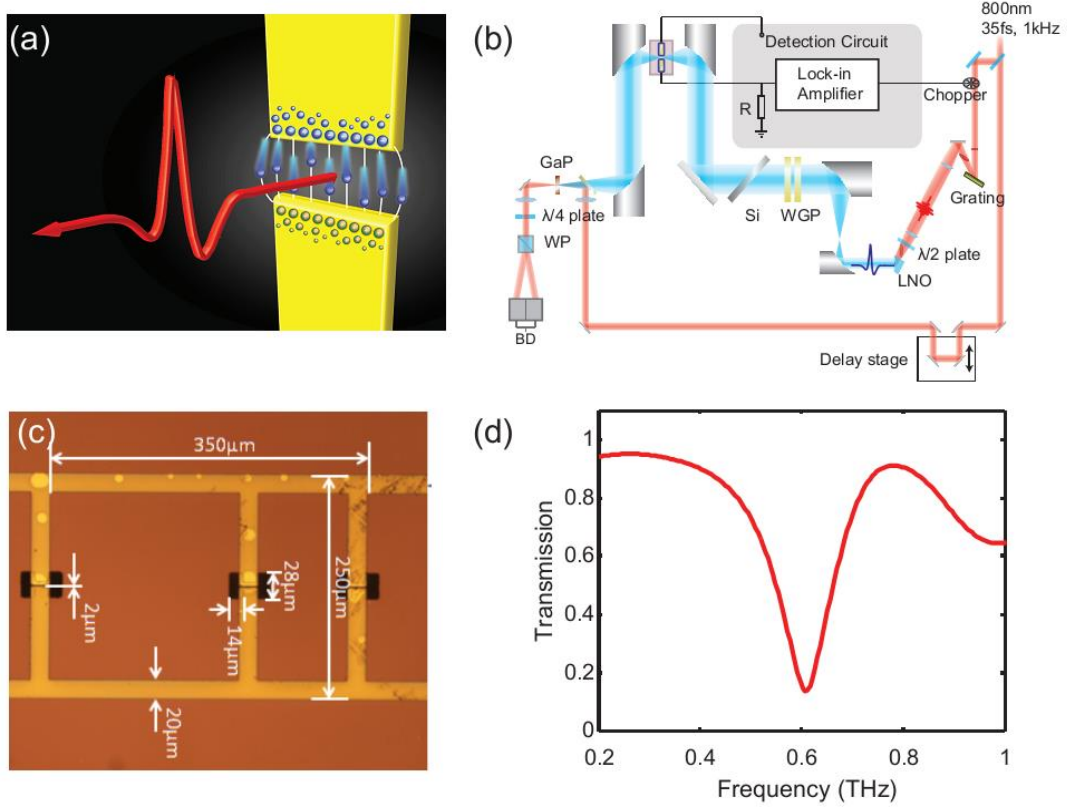

**Figure 1**

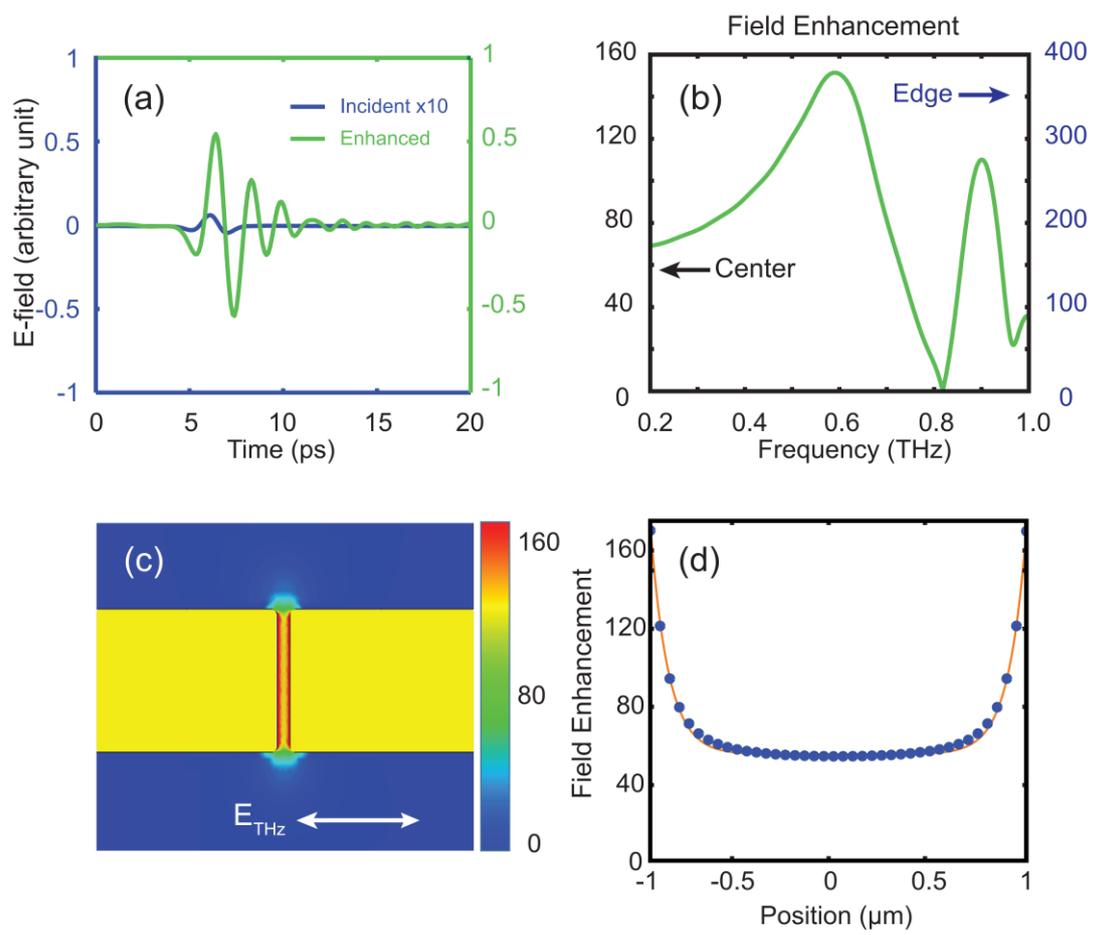

**Figure 2**

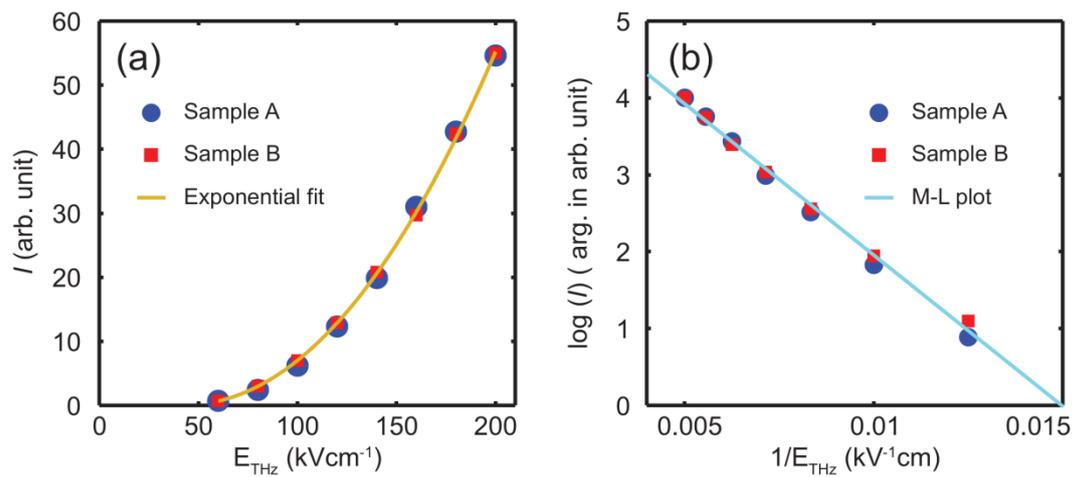

**Figure 3**

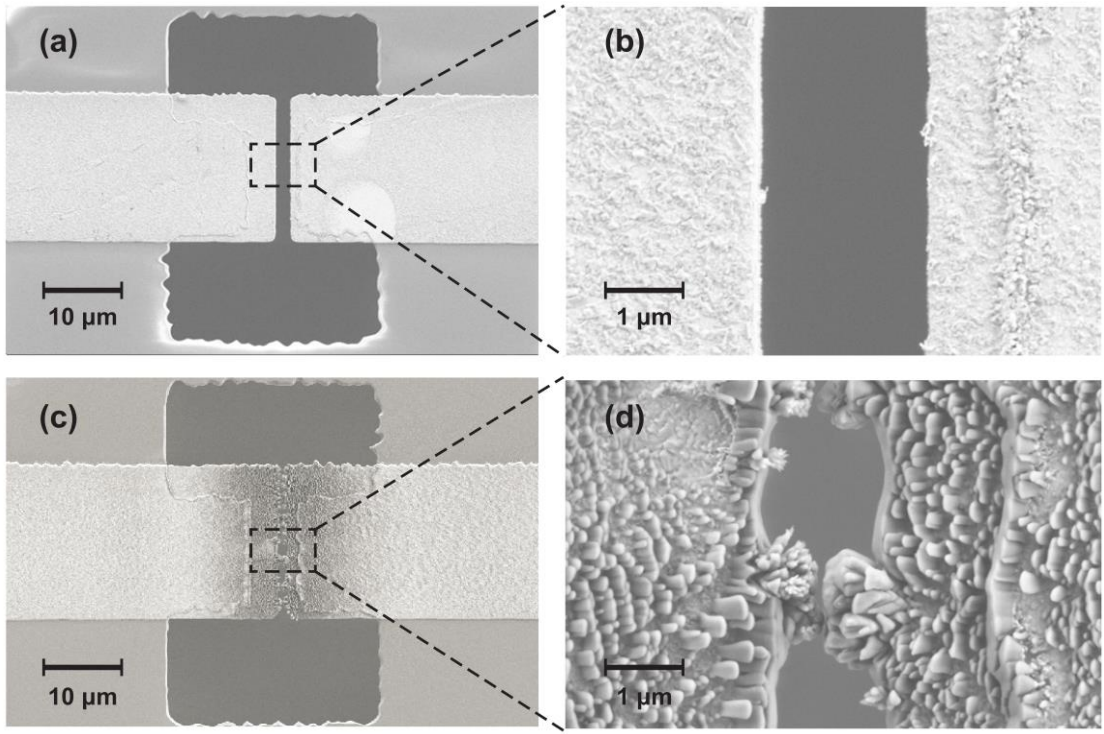

**Figure 4**

**Supplemental material:**

To simulate the in-gap acceleration of electrons emitted at the peak amplitude of the THz pulse, we use the *non-relativistic* equation of motion for a free electron in an AC field.

$$m_e \ddot{x} = -ef(x)E(t)$$

$$f(x) = 108\left[\exp\left(-\frac{x}{l_f}\right) + \exp\left(\frac{x-d}{l_f}\right)\right] + 60, x \in [0, 2\ \mu m]$$

$$E(t) = E_0 \exp\left(-2ln2\frac{t^2}{\tau^2}\right)\cos(2\pi\nu t)$$

Initial conditions: $x_{t=0} = 0; \dot{x}_{t=0} = 0$,

where $m_e$ is the electron rest mass, $e$ elementary charge, $f(x)$ is the spatial distribution of field enhancement, $l_f$ is the decay length of in-gap THz field (~0.094 μm), $\tau$ is the pulse duration of incident THz pulse (~1 ps), and $\nu$ is the central carrier frequency (~1 THz).

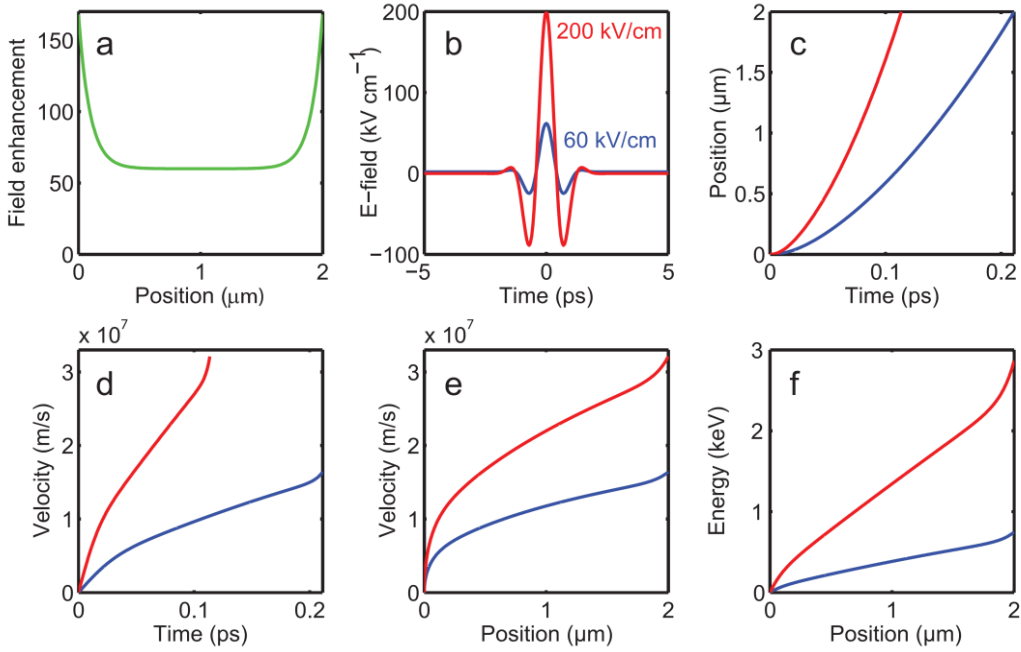

**Figure 1S**

Figure 1S (a) (b) shows the spatial distribution of field enhancement of *f(x)* and time-domain THz pulse *E(t)*, respectively. For the in-gap current simulations, we use a THz pulse (*E(t)*) that peaks at t=0. That a free electron is located at the gold-vacuum interface at t=0 that is subsequently accelerated by the in-gap E-field.

Figure 1S (c) (d) show the displacement and velocity of the free electron as a function of time. The time scale for electron to complete its flight is 0.1 to 2 ps. This verifies the **sub-cycle** nature of the emission process. One also notices that the calculated velocity is over $3*10^7$ m/s, which is a tenth of speed of light. Recall that the calculation is done at incident field amplitude of 200 kV/cm. It is likely that the accelerated electron has entered **relativistic regime** under THz field (200 kV/cm). This is because of the uniqueness of the THz wave, i.e. high amplitude, low frequency, which is difficult to achieve in other experiments using mid-IR and near-IR excitation. Figure 1S (e) (f) shows the velocity and kinetic energy of emitted electron as a function of its displacement.

In general, electron emission from a metallic tip/gap is a two-step process: (1) Ionization through MPI (Multi-photon ionization), ATI (Above threshold ionization) or TI (Tunneling ionization); (2) Acceleration by the oscillating electric field. From the perspective of maximum electron acceleration, it is advantageous to accelerate the electron in a strong and slow-varying field, such that electrons can utilize the full energy of the electro-magnetic field to gain maximum forward momentum while suppressing the quiver motion. As such, THz pulses push electrons into high momentum regime more easily than higher frequency fields.